\documentclass{ws-ijmpa}
\usepackage[super,compress]{cite}
\usepackage{color,graphicx}
\usepackage[usenames,dvipsnames,svgnames,table]{xcolor}
\usepackage{booktabs}

\usepackage[ps2pdf,
            colorlinks=true,
            linkcolor=blue,
            urlcolor=blue,
            citecolor=green,          
            bookmarks=true,
            bookmarksnumbered=true,
            breaklinks=true,
            pdfpagemode=Fullscreen,
            pdfstartview=FitBH]{hyperref}

\definecolor{gesfpurple}{rgb}{0.47,0.19,0.42}
\newcommand{\gpurple}[1]{{\color{gesfpurple} #1}}
\definecolor{gesflanse}{rgb}{0.00,0.50,0.50}

\definecolor{gesfblue}{rgb}{0.08,0.42,0.76}
\newcommand{\gblue}[1]{{\color{gesfblue} #1}}
\definecolor{gesfred}{rgb}{1,0,0}
\newcommand{\gred}[1]{{\color{gesfred} #1}}
\newcommand{\ured}[1]{{\color{gesfred}\underline{#1}}}
\definecolor{gesfwhite}{rgb}{1,1,1}

\definecolor{gesfblack}{rgb}{0,0,0}

\newcommand{\gsec}[1]{{\hypersetup{linkcolor=red}Sec.~\ref{#1}\hypersetup{linkcolor=blue}}}

\newcommand{\geqn}[1]{\hypersetup{linkcolor=blue}(\ref{#1})\hypersetup{linkcolor=blue}}
\newcommand{\gfig}[1]{{\hypersetup{linkcolor=violet}Fig.~\ref{#1}\hypersetup{linkcolor=blue}}}
\newcommand{\gtab}[1]{{\hypersetup{linkcolor=gesflanse}Tab.~\ref{#1}\hypersetup{linkcolor=blue}}}

\begin{document}
\markboth{Shao-Feng Ge, Hong-Jian He, Rui-Qing Xiao}{Testing Higgs Coupling Precision and New Physics Scales at Lepton Colliders}

\catchline{}{}{}{}{}
%
%%%%%%%%%%%%%%%%%%%%%%%%%%%%%%%%%%%%%%%%%%%%%%%%%%%%%%%%%%%%%%%%%%%%

\title{\large Testing Higgs Coupling Precision and New Physics Scales \\[1mm] at Lepton Colliders}

\author{Shao-Feng Ge \footnote{Presenter of the talk ``{\it New Physics Scales to be Probed at Lepton Colliders}'' at the IAS Program on High Energy Physics on January 11, 2016.}}
\address{Max-Planck-Institut f\"{u}r Kernphysik, Heidelberg, Germany\\
gesf02@gmail.com}

\author{Hong-Jian He}
\address{Institute of Modern Physics and Center for High Energy Physics,\\
                Tsinghua University, Beijing 100084, China \\
         Center for High Energy Physics, Peking University, Beijing 100871, China \\
         hjhe@tsinghua.edu.cn}

\author{Rui-Qing Xiao}
\address{Institute of Modern Physics and Center for High Energy Physics,\\
                Tsinghua University, Beijing 100084, China \\
         Lawrence Berkeley National Laboratory, Berkeley, California 94720, USA \\
         ruiqingxiao@lbl.gov}

\maketitle

\begin{history}
\received{Day Month Year}
\revised{\today}
\end{history}

\begin{abstract}

The next-generation lepton colliders, such as CEPC, FCC-ee, and ILC
will make precision measurement of the Higgs boson properties. We first
extract the Higgs coupling precision from Higgs observables at CEPC to
illustrate the potential of future lepton colliders. Depending on the
related event rates, the precision can reach percentage level for most
couplings. Then, we try to estimate the new physics scales that can be
indirectly probed with Higgs and electroweak precision observables.
The Higgs observables, together with the existing
electroweak precision observables, can probe new physics up to 10\,TeV
(40\,TeV for the gluon-related operator $\mathcal O_g$) at 95\% C.L.
Including the $Z/W$ mass measurements and $Z$-pole observables at CEPC
further pushes the limit up to 35\,TeV. Although $Z$-pole running is originally
for the purpose of machine calibration, it can be as important as the Higgs 
observables for probing the new physics scales indirectly. The indirect
probe of new physics scales at lepton colliders can mainly cover the
energy range to be explored by the following hadron colliders of
$pp (50-100$\,TeV), such as SPPC and FCC-hh.

\end{abstract}

\ccode{PACS numbers:}

%\tableofcontents

\section{Introduction}

With the discovery of Higgs boson \cite{Higgs} at LHC \cite{Higgs12},
the spectrum of the SM has been completed.
This culminates in the success of searches that lasted for decades \cite{HiggsProfile}.
Nevertheless, there are already many motivations for making precision
measurement of the Higgs coupling and testing the new physics beyond the SM.

The Higgs boson discovery is usually regarded as a big success of particle physics, which
seems to have completed the SM. Nevertheless, the SM not only requires the
existence of Higgs boson, but also dictates its interactions. To fully test the SM, it is necessary to measure all interactions that the Higgs boson
participates.
Although the spectrum of the SM has already been completed, the SM itself still
requires further experimental test. In this sense, the name ``particle physics''
is somewhat misleading with over-emphasis on
its particle content, and we must not forget {\it all the interaction forces} between these particles. Even within the SM, we are well motivated to test
the properties of the Higgs boson.

The existence of such a $125\,$GeV scalar is truly profound. It can provide
masses to massive gauge bosons and fermions. Its coupling constants with the SM
fermions and gauge bosons will be measured by the LHC to about $10\% - 20\%$ precisions as compared to the SM predictions \cite{LHC-fit}.
In this sense, we now understand how the mechanism of mass
generation can happen with a single vacuum expectation value (VEV). Nevertheless, 
how the Higgs acquires nonzero VEV has deep connection with the vacuum stability and
Higgs inflation \cite{diphoton}, but has not been tested experimentally yet
\cite{pp-h3}.
Especially, in the SM the Higgs mass receives
quadratic divergence from loop corrections and hence becomes radiatively
unnatural if the SM is valid up to very high energy scale. The Higgs mass
is radiatively unnatural. In addition, the Yukawa couplings span several orders
of magnitude and hence is hierarchically unnatural.
A satisfactory model shall make this hierarchical
unnaturalness in Yukawa couplings understandable \footnote{Related to fermion masses,
the quark and neutrino mixings have been experimentally established but we
are still not sure how to explain. Is the mixing pattern just coincidence
or consequence of some flavor symmetry? If there is any flavor symmetry
dictating the mixing pattern, what is it? Is the flavor symmetry discrete
or continuous? Especially, we shall keep in
mind that flavor symmetry has to be broken and the mixing pattern
can be determined by residual symmetry that can survive the electroweak
symmetry breaking rather than the full flavor
symmetry imposed on the fundamental Lagrangian \cite{residual}.}.

The next-generation lepton colliders, including CEPC \cite{CEPC},
FCC-ee \cite{FCC-ee}, and ILC \cite{ILC},
are designed for making precision measurement of the Higgs properties.
All three candidate machines can run at $\sqrt s = 250\,$GeV as Higgs factory
by producing Higgs boson through Higgsstrahlung $e^+ e^- \rightarrow Zh$
and $WW$ fusion $e^+ e^- \rightarrow \nu \bar \nu h$ processes. The CEPC
with $5\,\mbox{ab}^{-1}$ of integrated luminosity can roughly produce $10^6$
Higgs bosons. From a naive estimation, the statistical fluctuation can
reach $\mathcal O(0.1\%)$ level for inclusive observables. Considering
the fact that the Higgs can decay via various channels, the precision on
its coupling can typically reach $\mathcal O(1\%)$ level.

In this talk, we first summarize in \gsec{sec:precision} the Higgs coupling
precision that can be reached at CEPC based on the assumption that the
Higgs production and decay processes can be described by rescaling the
SM predictions. We then use the Higgs observables (including both production
and decay rates of the Higgs boson), $M_Z/M_W$ mass measurements, and
$Z$-pole observables to estimate the new physics scales via dimension-6
operators in \gsec{sec:dim6}.
Finally, our conclusion can be found in \gsec{sec:conclusion}.
Interested readers can check our full paper \cite{dim6} for details.

\section{Higgs Coupling Precision}
\label{sec:precision}

Of the $10^6$ Higgs bosons, most of them are produced through the Higgsstrahlung
process, $e^+ e^- \rightarrow Z h$. Since there are only two particles
in the final state and the initial state is well defined, the Higgs boson
can be reconstructed from the $Z$ boson, $p_h = - p_Z$. This is the so-called
{\it recoil mass reconstruction technique} \cite{recoil} which allows
inclusive measurement of the Higgsstrahlung cross section $\sigma(Zh)$. Among the
selected events, the Higgs decay rate $\sigma(Zh) \times \mbox{Br}(h \rightarrow ii)$
of various channels can be measured independently. The ratio of the decay
and production rates is the corresponding decay branching ratio
$\mbox{Br}(h \rightarrow ii)$. In this way, the Higgs decay branching ratios can
be measured in a model-independent way at lepton colliders.
For the $WW$ fusion process, $e^+ e^- \rightarrow \nu \bar \nu h$, only the
$h \rightarrow b b$ channel has large enough rate $\sigma(\nu \bar \nu h)
\times \mbox{Br}(h \rightarrow b b)$. Together with the decay
branching ratio $\mbox{Br}(h \rightarrow bb)$ inferred from Higgsstrahlung,
the cross section $\sigma(\nu \bar \nu h)$ can also be extracted as the ratio between
the directly measured $\sigma(\nu \bar \nu h) \times \mbox{Br}(h \rightarrow bb)$
and the already determined branching ratio $\mbox{Br}(h \rightarrow bb)$.
In \gtab{tab:inputs}, we summarize the direct Higgs observables in black boxes
and label the inputs to $\chi^2$ fit in red color.

\begin{table}[h]
\centering
\tbl{The estimated $1\sigma$ precision of Higgs observables at CEPC \cite{CEPC,CEPC16}.
The quantities in black box are experimental observables that can be directly
measured while the quantities labeled in red color are the inputs to our $\chi^2$
fit analysis.~~~~~~~~~~~~~~~~~~~~~~~~~}
{\begin{tabular}{clccc}
  ${\bf \boldsymbol \Delta M_h}$        &    ${\bf \boldsymbol \Gamma_h}$     & \fbox{$\gred{\bf \boldsymbol \sigma(Zh)}$}  &
\fbox{$\gred{\bf \boldsymbol \sigma(\boldsymbol \nu \bar{\boldsymbol \nu} h)} \times\mathrm{Br}(h\to bb)$}     \\
\hline
     2.6 MeV          &       2.8\%            &   0.5\%         &     2.8\%       \\
\\
& Decay Mode  &    \fbox{$\sigma(Zh)\times\mathrm{Br}$}      &   $\gred{\bf \mathrm{Br}}$   \\ \hline
& $h\to bb$           &    0.21\%      &      0.54\%	        \\
& $h\to cc$           &    2.5\%       &      2.5\%	        \\
& $h\to gg$           &    1.7\%       &      1.8\%	      	\\
& $h\to \tau\tau$     &    1.2\%       &      1.3\%	       	\\
& $h\to WW$           &    1.4\%       &      1.5\%	       	\\
& $h\to ZZ$           &    4.3\%       &      4.3\%	        \\
& $h\to\gamma\gamma$  &	   9.0\%       &      9.0\%	       	\\
& $h\to\mu\mu$        &    17\%        &       17\%	       	\\
& $h\to {\rm invisible}$    &    --      &      0.14\% 
\end{tabular}}
\label{tab:inputs}
\end{table}

To extract the precision on the Higgs coupling with the SM particles, we rescale the
SM prediction, $g_{hii} / g^{\rm sm}_{hii} \equiv \kappa_i$, and parametrize
the deviation from the SM as $\delta \kappa_i \equiv \kappa_i - 1$. The Higgsstrahlung
and $WW$ fusion cross sections are then modulated by the Higgs couplings with
$Z$ and $W$ bosons,
\begin{equation}
  \frac {\delta \sigma (Zh)}{\sigma(Zh)}
=
  \kappa^2_Z - 1
\simeq
  2 \delta \kappa_Z \,,
\qquad
  \frac {\delta \sigma (\nu \bar \nu h)}{\sigma(\nu \bar \nu h)}
=
  \kappa^2_W - 1
\simeq
  2 \delta \kappa_W \,,
\end{equation}
respectively. Similarly, the decay widths are also modulated by the corresponding
rescaling $\kappa$ factors,
\begin{equation}
  \frac {\delta \Gamma_{hii}}{\Gamma^{\rm sm}_{hii}}
=
  \kappa^2_i - 1
\simeq
  2 \delta \kappa_i \,,
\qquad
  \frac {\Gamma_{inv}}{\Gamma^{\rm sm}_{tot}}
=
  \mbox{Br}(h \rightarrow invisibles)
\equiv
  \delta \kappa_{inv} \,.
\label{eq:Gamma}
\end{equation}
Since no invisible decay mode is present in the SM, its contribution is
parametrized directly as $\delta \kappa_{inv}$ which vanishes when the SM is
recovered. In contrast, for those channels already existing in the SM, the
deviation $\delta \kappa_i$ is the difference from $1$.
The total decay width is then the sum over all decay
channels, $\Gamma^{\rm sm}_{tot} \equiv \sum_i \Gamma^{\rm sm}_{hii}$. 

To fit the quantities labeled as red color in \gtab{tab:inputs}, the rescaled
decay widths need to be expressed as decay branching ratios,
\begin{equation}
  \mbox{Br}^{\rm th}_i
\simeq
  \mbox{Br}^{\rm sm}_i
\left[
  1
+ \left( 1 - \mbox{Br}^{\rm sm}_i \right) \frac{\delta\Gamma_i}{\Gamma_i}
- \sum_{j \neq i} \mbox{Br}^{\rm sm}_j \frac{\delta \Gamma_j}{\Gamma_j}
\right] \,,
\label{eq:Br-i}
\end{equation}
where
$\mbox{Br}(h \rightarrow ii) \equiv \Gamma_{hii} / \Gamma_{tot}$ and
$\Gamma_{tot} \equiv \sum_i \Gamma_{hii}$.
Note that the rescaling is no longer an overall factor. For the parametrization
of both channels existing in the SM and the invisible channel in \geqn{eq:Gamma},
the theoretical prediction can be expanded as,
\begin{equation}
  \mbox{Br}^{\rm th}_i
\simeq
  \mbox{Br}^{\rm sm}_i
\left(
  1
+ \sum_j A_{ij} \delta \kappa_j
\right) ,
\qquad
  \mbox{Br}^{\rm th}_{inv}
\simeq
  \delta \kappa_{inv} ,
\label{eq:BR-expanded}
\end{equation}
where the coefficient matrix $A$ has elements,
\begin{equation}
  A_{ij}
=
  2 ( \delta_{ij} - \mbox{Br}^{\rm sm}_j ) ,
\qquad
  A_{i, inv}^{} =  - 1 ,
\qquad
  A_{inv, i}^{} =   0 ,
\qquad
  A_{inv, inv}^{} =  1 .
\label{eq:A}
\end{equation}
In the decay branching ratios, all decay channels can affect each other.
For the channel with larger branching ratio $\mbox{Br}^{\rm sm}_i$ in the SM,
it can have larger effect ($A_{ji} = - \mbox{Br}^{\rm sm}_i$) on the others
but smaller effect ($A_{ii} = 1 - \mbox{Br}^{\rm sm}_i$) on itself. The only
exception is the invisible channel which has effect of equal size
($A_{inv, inv} = 1$ for its own and $A_{i, inv} = -1$ for the others)
on all channels. Note that the invisible channel can be affected only by itself.

As summarized in \gtab{tab:inputs}, 9 decay channels from 2 production modes
can achieve reasonable precision. To keep the fit as general as possible
when estimating the precision on measuring the deviation of Higgs couplings from
the SM prediction, 9 scaling $\kappa_i$ ($i = b, c, g, \tau, W, Z, \gamma, \mu, inv$)
are introduced for the nine decay channels, respectively. Note that in the SM, the
Higgs decay into a pair of photons or gluons is induced by triangle loops with
fermion or $W$ boson that can directly couple to Higgs and hence is not fully
independent. Nevertheless, independent scaling factors $\kappa_\gamma$ and
$\kappa_g$ are assigned to the $h \rightarrow \gamma \gamma$ and $h \rightarrow gg$
decay widths for generality. The 9-parameter fit
is based on SM and can also accommodate new physics contributions.
\begin{table}[h]
\centering
\tbl{The $1\sigma$ precisions on measuring Higgs couplings
         at CEPC (250GeV, 5ab$^{-1}$), in comparison with
         LHC (14TeV, 300fb$^{-1}$), HL-LHC (14TeV, 3ab$^{-1}$) and
         ILC (250GeV, 250fb$^{-1}$) + (500GeV, 500fb$^{-1}$).
         The numbers for LHC, HL-LHC, and ILC are obtained from \cite{Peskin}.}
{
\begin{tabular}{c|cc|cc|c}
 Precision (\%) & \multicolumn{2}{c|}{CEPC} & LHC & HL-LHC & ILC-250+500 \\
\hline
 $\kappa_Z$      & \gpurple{0.249} & \gpurple{0.249} &  8.5 & 6.3 & 0.50 \\
 $\kappa_W$      &  1.21 &  1.21 &  5.4 & 3.3 & \gpurple{0.46} \\
 $\kappa_\gamma$ &  \gpurple{4.67} &  \gpurple{4.67} &  9.0 & 6.5 & 8.6  \\
 $\kappa_g$      &  1.55 &  1.55 &  6.9 & 4.8 & 2.0  \\
 $\kappa_b$      &  1.28 &  1.28 & 14.9 & 8.5 & \gpurple{0.97} \\
 $\kappa_c$      &  1.76 &  1.76 & --   & --  & 2.6  \\
 $\kappa_\tau$   &  1.39 &  1.39 &  9.5 & 6.5 & 2.0  \\
 $\kappa_\mu$    &  --   &  \gpurple{8.59} & --   & --  & --   \\
 $\rm Br_{inv}$  & 0.135 & 0.135 &  8.0 & 4.0 & 0.52 \\
 $\Gamma_h$      & 2.8   & 2.8   & --   & --  & --
\end{tabular}}
\label{tab:precision}
\end{table}

Using the technique of analytical $\chi^2$ fit \cite{dim6,JUNO} delivered in
the BSMfitter package \cite{BSMfitter}, we estimate the precision on Higgs couplings
and summarize the results in \gtab{tab:precision}. 
\begin{itemize}
\item Roughly speaking, the uncertainty
is mainly determined by statistical fluctuations and hence the SM prediction of decay
branching ratios $\mbox{Br}^{\rm sm}_i$. 

\item Nevertheless, the precision on $\kappa_Z$ is
much better than the precision on $\kappa_W$ although the Higgs decay $h \rightarrow WW$
($\mbox{Br}^{\rm sm}_{h \rightarrow WW} = 22.5\%$) has larger branching ratio than
$h \rightarrow ZZ$ ($\mbox{Br}^{\rm sm}_{h \rightarrow ZZ} = 2.77\%$). The
additional constraint comes from the Higgsstrahlung cross section $\sigma(Zh)$
which is an inclusive observable and hence has the largest event rate. Similarly,
$\kappa_W$ also receives constraint from both decay branching ratios and the
$WW$ fusion cross section $\sigma(\nu \bar \nu h)$.

\item Apart from these, the others
are only constrained by decay branching ratios. Of all rescaling factors, $\kappa_\mu$
has the worst precision since $h \rightarrow \mu \mu$
($\mbox{Br}^{\rm sm}_{h \rightarrow \mu \mu} = 0.023\%$) has the least number of
events and hence the worst statistics. For comparison, we show two fits with or
without $\kappa_\mu$ and find that all other numbers are not affect. This demonstrates
what we argued that a channel $h \rightarrow ii$ can affect other channels with weight
$\mbox{Br}^{\rm sm}_i$. For $h \rightarrow \mu \mu$, its branching ratio 
$\mbox{Br}^{\rm sm}_{h \rightarrow \mu \mu} \simeq 0.023\%$ is negligibly small.

\item 
Although the branching ratio of
$h \rightarrow \gamma \gamma$ is only around $1\%$ of $h \rightarrow WW$ and hence
has 10 times larger uncertainty from naive estimation of statistical fluctuation,
the precision on $\kappa_\gamma$ is not that worse than $\kappa_W$. This is
because that photon can be measured much better than the $W$ boson. The photon
can be probed directly but the $W$ boson needs to first decay. 

\item Note that
the invisible decay has the best uncertainty as shown in \gtab{tab:inputs}, rendering
$\mbox{Br}_{inv} \simeq \delta \kappa_{inv}$ to be mainly determined by this
single channel $h \rightarrow invisible$. We can see that the fitted precision
$0.135\%$ in \gtab{tab:inputs} is roughly the same number as the original value
$0.14\%$ in \gtab{tab:precision} since the coefficient $A_{inv, inv}$ in
\geqn{eq:A} is 1. 

\item Finally, the combined precision on the total decay width
$\Gamma_h$ is simply the value directly from detector simulation in \gtab{tab:inputs}.
This is because $\Gamma_h$ is an independent variable and not entangled with the
rescaling factors $\kappa_i$.
\end{itemize}
 
\begin{figure}[h]
\centering
\includegraphics[width=0.95\textwidth]{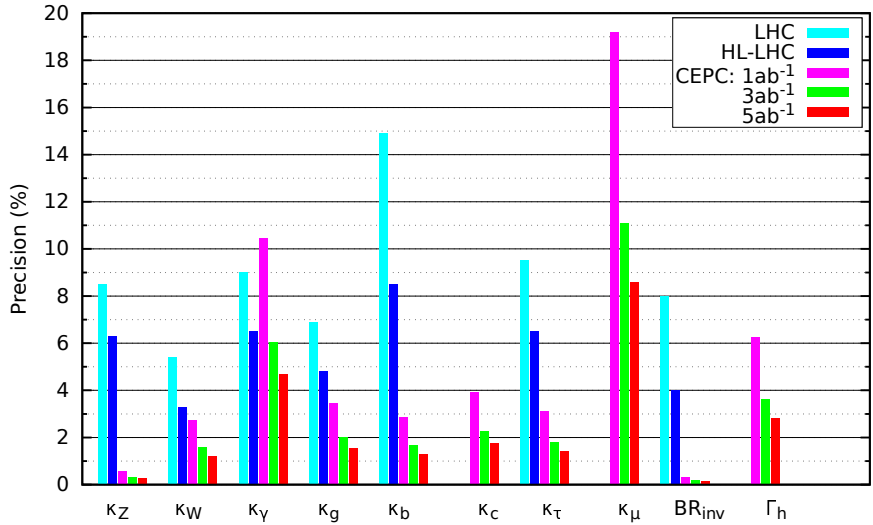}
\caption{The $1 \sigma$ precision on the Higgs couplings at CEPC (250\,GeV)
         with integrated luminosity $(1,3,5)\,\mbox{ab}^{-1}$, respectively, in
         comparison with the results at LHC (14\,TeV, 300\,fb$^{-1}$) and
         HL-LHC (14\,TeV, 3\,ab$^{-1}$).}
\label{fig:precision}
\end{figure}
For comparison, we also show in \gtab{tab:precision} the precision at LHC, HL-LHC,
and ILC \cite{Peskin}. At hadron colliders like LHC and HL-LHC, not all channels
can be measured and the precision is usually much worse than at lepton colliders.
To make better presentation, the results are depicted in \gfig{fig:precision}.
The reachable precision at CEPC can receive significant improvement from the
expected measurement at LHC and HL-LHC. The largest difference appears in
$\kappa_Z$ and $\mbox{Br}_{inv}$ with improvement more than one order of magnitude.
In addition, hadron colliders can not measure the Higgs decay width due to large
energy uncertainty.
For ILC, it can make similar measurements as
CEPC. Nevertheless, for most channels the precision is not as good as at CEPC due
to smaller integrated luminosity. There are two exceptions, one is $\kappa_W$
and the other is $\kappa_b$. For $\kappa_W$, the enhancement comes from the
running at $\sqrt s = 500$\,GeV where the $WW$ production of Higgs can be
significantly enhanced to introduce better constraint on
$\sigma(\nu \bar \nu h) = \kappa^2_W \sigma^{\rm sm}(\nu \bar \nu h)$.

\section{New Physics Scales}
\label{sec:dim6}

The purpose of building lepton colliders is not just for precision measurement
of the Higgs couplings. Our final goal is new physics beyond the SM. Nevertheless,
the energy of lepton colliders is probably not enough to produce the new particles
which are usually expected to be heavy. Since the Higgs boson is newly discovered
and other SM particles have already been precisely measured to some extent, if
there is any new physics, it has large chance to appear in the Higgs couplings.
It has large chance for new physics to be measured at lepton colliders indirectly.

The effect of new physics from higher energy can be parametrized in terms of
dimension-6 operators. On the basis of the SM, the Lagrangian is extended by adding
effective operators,
\begin{equation}
  \mathcal L
=
  \mathcal L_{SM}
+
  \sum_{ij} \frac {y_{ij} \sim \mathcal O(1)}{\Lambda \sim 10^{14} \mbox{GeV}}
  \left( \overline L_i \widetilde H \right) 
  \left( \widetilde H^\dagger L_j \right) 
+
  \sum_i \frac {c_i}{\Lambda^2} \mathcal O_i \,,
\end{equation}
where $\widetilde H \equiv \tau_2 H^*$ is the CP conjugate of the Higgs doublet $H$.
For completeness, we also show the dimension-5 operators which can provide neutrino
mass matrix. For $y_{ij} \sim \mathcal O(1)$, the tiny neutrino masses can be
produced if the cutoff $\Lambda \sim 10^{14}$\,GeV is at the GUT scale. Note that
the dimension-5 operators have lower suppression than the dimension-6 operators
and hence are expected to have larger effect on low-energy physics. This is probably
the reason why we have already measured neutrino mass and mixing in neutrino
oscillation experiments but have not seen the dimension-6 operators.

\begin{table}[h]
\fontsize{8}{10}
\setlength{\tabcolsep}{0.7mm}
\centering
\tbl{The CP-even dimension-6 operators related to Higgs and electroweak precision
     observables to be measured at lepton colliders.}
{
\begin{tabular}{ccc}
  Higgs & EW Gauge Bosons & Fermions
\\
\hline
  $\gblue{\bf \boldsymbol {\mathcal O}_H^{}}\! = \frac 1 2 (\partial_\mu |\gpurple{\bf H}|^2)^2$
& $\gblue{\bf \boldsymbol {\mathcal O}_{WW}^{}}\! = g^2 |\gpurple{\bf H}|^2 W^a_{\mu \nu} W^{a\mu \nu}$
& $\gblue{\bf \boldsymbol {\mathcal O}^{(3)}_L}\! = (i \gpurple{\bf H}^\dagger\sigma^a\!\! \stackrel \leftrightarrow D_\mu^{}\!\! \gpurple{\bf H})
(\overline \Psi_L^{} \gamma^\mu \sigma^a \Psi_L^{})$
\\
  $\gblue{\bf \boldsymbol {\mathcal O}_T^{}}\! = \frac 1 2 (\gpurple{\bf H}^\dagger\! \stackrel\leftrightarrow D_\mu^{}\! \gpurple{\bf H})^2$
& $\gblue{\bf \boldsymbol {\mathcal O}_{BB}^{}}\! = g^2 |\gpurple{\bf H}|^2 B_{\mu \nu} B^{\mu \nu}$
& $\gblue{\bf \boldsymbol {\mathcal O}^{(3)}_{LL}}\! =
  (\overline \Psi_L^{} \gamma_\mu \sigma^a \Psi_L^{})
  (\overline \Psi_L^{} \gamma^\mu \sigma^a \Psi_L^{})$
\\
& $\gblue{\bf \boldsymbol {\mathcal O}_{WB}^{}}\! = g g' \gpurple{\bf H}^\dagger \sigma^a \gpurple{\bf H} W^a_{\mu \nu} B^{\mu \nu}$
& $\gblue{\bf \boldsymbol {\mathcal O}_L^{}}\! = (i \gpurple{\bf H}^\dagger\! \stackrel \leftrightarrow D_\mu^{}\! \gpurple{\bf H})
(\overline \Psi_L^{} \gamma^\mu \Psi_L^{})$
\\
  Gluon
& ~$\gblue{\bf \boldsymbol {\mathcal O}_{HW}^{}}\! = i g (D^\mu \gpurple{\bf H})^\dagger \sigma^a (D^\nu \gpurple{\bf H}) W^a_{\mu \nu}$~
& $\gblue{\bf \boldsymbol {\mathcal O}_R^{}}\! = (i \gpurple{\bf H}^\dagger\!\! \stackrel \leftrightarrow D_\mu^{}\!\! \gpurple{\bf H})
  (\overline \psi_R^{} \gamma^\mu \psi_R^{})$
\\\cmidrule{1-1}
  $\gblue{\bf \boldsymbol {\mathcal O}_g^{}}\! = g_s^2 |\gpurple{\bf H}|^2 G^a_{\mu \nu} G^{a\mu \nu}$
& $\gblue{\bf \boldsymbol {\mathcal O}_{HB}^{}}\! = i g' (D^\mu \gpurple{\bf H})^\dagger (D^\nu \gpurple{\bf H}) B_{\mu \nu}$
&
\end{tabular}}
\label{tab:O}
\end{table}

Since the Higgs couplings are the thing to be measured at lepton colliders, we
list all related CP-even operators in \gtab{tab:O}. All operators involve the
Higgs doublet $H$ with the only exception of $\mathcal O^{(3)}_{LL}$ which
is a four-fermion operator and only affects the Fermi constant $G_F$. In this
set, all operators are independent by removing redundant operators like
$\mathcal O_W$ and $\mathcal O_B$. For simplicity, we will not elaborate the
details of how these operators can modify the SM prediction of Higgs and
electroweak precision observables \cite{dim6} and only focus on the physical
consequences.

\begin{table}[h]
\centering
\tbl{Inputs used to constrain the new physics scales of
     dimension-6 operators. The electroweak precision observables
     in the first four rows are taken from PDG2014 \cite{PDG14},
     and the $1\,\sigma$ precisions of Higgs measurements
     are taken from the CEPC detector simulations \cite{CEPC}.
     For the $WW$ fusion cross section $\sigma[\nu \bar \nu h]_{350\text{GeV}}$
     at $\sqrt{s}=350$GeV, we adopt the  
     TLEP estimate of its uncertainty\cite{FCC-ee} as an illustration.}
{\begin{tabular}{cccc}
Observables & Central Value & Relative Error & SM Prediction
\\
\hline
\gblue{$\boldsymbol \alpha$} & $7.297  352  569  8 \!\times\! 10^{-3}$ & $3.29 \!\times\! 10^{-10}$ & --
\\
\gblue{$\bf G_F$} & $1.166  378  7 \!\times\! 10^{-5} \mbox{GeV}^{-2}$ & $5.14 \!\times\! 10^{-7}$ & --
\\
\gblue{$\bf M_Z$} & 91.1876GeV & $2.3 \times 10^{-5}$ & --
\\
\gblue{$\bf M_W$} & 80.385GeV  & $1.87 \times 10^{-4}$ & --
\\
\hline
\gpurple{$\sigma[Zh]$} & -- & 0.51\% & -- \\
\gpurple{$\sigma[\nu \bar \nu h]$} & -- & 2.86\% & -- \\
\gpurple{{~~~~~~~~~$\sigma[\nu \bar \nu h]_{350\text{GeV}}$}} & -- & {0.75\%} & -- \\
\hline
\gpurple{$\text{Br}[WW]$} & -- & 1.6\% & 22.5\% \\
\gpurple{$\text{Br}[ZZ]$} & -- & 4.3\% & 2.77\% \\
\gpurple{$\text{Br}[bb]$} & -- & 0.57\% & 58.1\% \\
\gpurple{$\text{Br}[cc]$} & -- & 2.3\% & 2.10\% \\
\gpurple{$\text{Br}[gg]$} & -- & 1.7\% & 7.40\% \\
\gpurple{$\text{Br}[\tau \tau]$} & -- & 1.3\% & 6.64\% \\
\gpurple{$\text{Br}[\gamma \gamma]$} & -- & 9.0\% & 0.243\% \\
\gpurple{$\text{Br}[\mu \mu]$} & -- & 17\% & 0.023\%
\end{tabular}}
\label{tab:inputs2}
\end{table}

In \gtab{tab:inputs2}, we summarize the existing electroweak precision
observables \cite{PDG14} and the Higgs observables at CEPC \cite{CEPC}.
These observables are used to constraint the size of dimension-6 operators
with the following $\chi^2$ function,
\begin{equation}
  \chi^2 \left( \delta \alpha, \delta G_F, \delta M_Z, \frac {c_i^{}}{\Lambda^2} \right)
=
  \sum_j
\left[
  \frac{\mathcal O^{\rm th}_j \left( \delta \alpha, \delta G_F, \delta M_Z, \frac {c_i}{\Lambda^2} \right) - \mathcal O^{\rm exp}_j}
       {\Delta \mathcal O_j}
\right]^2 ,
\end{equation}
where $(\delta \alpha, \delta G_F, \delta M_Z)$ are shifts of the
corresponding parameter $(\alpha, G_F, M_Z)$ from their reference values,
\begin{subequations}
\begin{eqnarray}
  \alpha^{(\text{sm})}
& = &
  \alpha^{(r)} \left( 1 + \frac {\delta \alpha} \alpha \right) ,
\\
  G^{(\text{sm})}_F
& = &
  G^{(r)}_F \left( 1 + \frac {\delta G_F}{G_F} \right),
\\
  M^{(\text{sm})}_Z
& = &
  M^{(r)}_Z \left( 1 + \frac {\delta M_Z}{M_Z} \right).
\label{eq:Zscheme-inputs}
\end{eqnarray}
\end{subequations}
When doing $\chi^2$ fit, both the dimension-6 operator coefficient $c_j$ and 
parameter shifts $(\delta \alpha, \delta G_F, \delta M_Z)$ are treated as fitting
parameters on the equal footing. For convenience, we take the reference values
to be at the experimental central values,
$\alpha^{(r)} = 7.297  352  569  8 \times 10^{-3}$,
$G^{(r)}_F = 1.1663787 \times 10^{-5} \mbox{GeV}^{-2}$,
and $M^{(r)}_Z = 91.1876 \mbox{GeV}$, while keeping the shifts as small deviations.

This choice of fitting parameters is different from the conventional $Z$-scheme
by fixing the input parameters $(\alpha, G_F, M_Z)$ to their central values or
the $W$-scheme by fixing $(\alpha, M_Z, M_W)$ instead. In these scheme-dependent
approaches, only the central values of the input parameters are utilized while
their uncertainties are simply discarded. The scheme-dependent
approach is practically good enough if the input parameters are much more precise
than the other observables. This is the case for $Z$-scheme with current electroweak
precision measurements. As shown in \gtab{tab:inputs2}, the relative errors of
$(\alpha, G_F, M_Z)$ are negligibly small, at least one order of magnitude better
than $M_W$. Nevertheless, the situation changes when the uncertainty on $M_W$
is significantly improved to be comparable with the uncertainty on $M_Z$.
This is exactly the case for CEPC.

\subsection{Sensitivity Reach from Higgs Observables}

\begin{figure}[h]
\centering
\includegraphics[width=\textwidth]{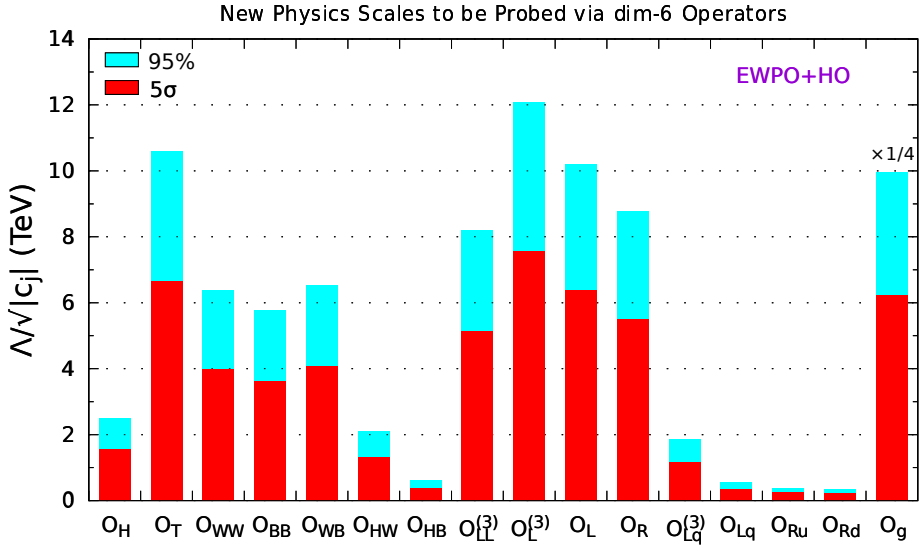}
\caption{The 95\% exclusion limits (blue) and $5\sigma$ discovery sensitivities (red) to
the new physics scales $\Lambda/\sqrt{|c_j|}$
by combining the current electroweak precision observables
($\alpha, G_F, M_Z, M_W$) \cite{PDG14} and the future Higgs observables
(Table\,\ref{tab:inputs2}) at the Higgs factory CEPC\,(250\,GeV) \cite{CEPC}
with a projected luminosity of 5\,ab$^{-1}$.
In the last column for $\mathcal O_g$, we have rescaled its height
by a factor $1/4$ to fit the plot, so its actual reach is
$\Lambda /\sqrt{|c_g|}=39.8\,$TeV.}
\label{fig:EW}
\end{figure}

Using all the observables in \gtab{tab:inputs2} (except
$\sigma[\nu \bar \nu h]_{350 \mbox{\tiny GeV}}$ which is taken from the TLEP estimation)
we first estimate the effect of Higgs observables on probing new physics indirectly
and show the results in \gfig{fig:EW}. The 95\% limit (blue) indicates the exclusion
sensitivity while the $5\sigma$ value (red) is the sensitivity for discovery. The following
discussions focus on the 95\% limit for simplicity.

\vspace{-3mm}
\begin{itemize}
\item We can see that the new physics
scales for $(\mathcal O_T, \mathcal O^{(3)}_L, \mathcal O_L)$ can be probed up
to as high as around 10\,TeV. Half of the operators
($\mathcal O_H$, $\mathcal O_{WW}$, $\mathcal O_{BB}$, $\mathcal O_{WB}$,
 $\mathcal O_{HW}$, $\mathcal O^{(3)}_{LL}$, $\mathcal O_R$, $\mathcal O^{(3)}_{Lq}$)
can be probed up to the scales $2 \sim 10$\,TeV which is already beyond the effective
scale that can be probed at LHC. The remaining ($\mathcal O_{HB}$,
$\mathcal O_{Lq}$, $\mathcal O_{Ru}$, $\mathcal O_{Rd}$) can only be probed below
1\,TeV. 

\item Note that the gluon involved operator $\mathcal O_g$ can even be probed
up to 40\,TeV already. The reason is that the Higgs decay into a pair of gluons,
$h \rightarrow gg$, is induced by triangle loops in the SM. In comparison, the
contribution of dimension-6 operators is also at the one-loop level and hence of
the same order as the SM prediction $\mbox{Br}^{\rm sm}_{h \rightarrow gg}$. 
Deviation in this channel then has magnified effect.

\end{itemize}

\subsection{The Improvement from Electroweak Precision Observables}

\begin{table}[h]
\centering
\tbl{Impacts of adding the current electroweak precision observables
($\alpha$, $G_F$, $M_Z$, $M_W$) \cite{PDG14}
on probing the new physics scales $\Lambda/\sqrt{|c_j|}$ (in TeV) at 95\% C.L.
The limits in the first row are
obtained from $\sigma(Zh)$ to be measured at the CEPC \cite{CEPC} only. The limits in
the second row are given by combining with the current $M_W$ measurement
plus $\sigma(Zh)$. Finally, the third row presents the limits by including
the current measurements of ($\alpha$, $G_F$, $M_Z$) altogether.
In the first two rows, ($\alpha$, $G_F$, $M_Z$) are fixed to their experimental central values
as in the $Z$-scheme, while the third row adopts the scheme-independent approach
by allowing all electroweak parameters to freely vary in each fit.
We label the entries of most significant improvements in red color with an underscore.}
{
\begin{tabular}{ccccccccccc}
  $\mathcal O_H^{}$ & $\mathcal O_T^{}$
& $\mathcal O_{WW}^{}$ & $\mathcal O_{BB}^{}$
& $\mathcal O_{WB}^{}$ & $\mathcal O_{HW}^{}$
& $\mathcal O_{HB}^{}$ & $\mathcal O^{(3)}_{LL}$
& $\mathcal O^{(3)}_L$ & $\mathcal O_L^{}$ & $\mathcal O_R^{}$
\\
\hline
 2.48 &       2.01  & 4.83 & 0.89  &       1.86  & 2.09 & 0.567 &       5.38  & 11.6 & 10.2 & 8.78 \\
 2.48 & \ured{10.6} & 4.83 & 0.89  & \ured{5.16} & 2.09 & 0.567 & \ured{8.22} & 12.1 & 10.2 & 8.78 \\
 2.48 &       10.6  & 4.83 & 0.875 &       5.12  & 2.09 & 0.567 &       8.15  & 12.1 & 10.2 & 8.78
\end{tabular}}
\label{tab:scheme}
\end{table}
Among the most sensitive operators, the limit for $\mathcal O_T$ is mainly
provided by $M_W$. In \gtab{tab:scheme} we show the effect of imposing the
existing electroweak precision observables. The first row shows the effect
of only $\sigma(Zh)$. On this basis, $M_W$ is added in the second row
and then all of ($\alpha$, $G_F$, $M_Z$, $M_W$) in the third row.
We can see that the most significant enhancement from $M_W$ appears
in $\mathcal O_T$ by promoting its limit from
2\,TeV up to 10\,TeV. In addition, the limits on $\mathcal O_{WB}$ and
$\mathcal O^{(3)}_{LL}$ can also be increased by a factor of $2 \sim 3$.
Nevertheless, further imposing the measurements of ($\alpha$, $G_F$, $M_Z$)
has no effective help since their effect is basically fixing the electroweak
parameters ($g$, $g'$, $v$).

\begin{table}[h]
\vspace{-3mm}
\centering
\tbl{The projected precision ($1\sigma$) of $Z$ and $W$ mass measurements
         to be achieved at the CEPC \cite{CEPC,Zpole}.}
{
\begin{tabular}{ccc}
~Observables~ & Relative Error & ~Absolute Error~ \\
\hline
$M_Z^{}$ & $(0.55  - 1.1) \!\times\! 10^{-5}$ & $(0.5 - 1)$\,MeV \\
$M_W^{}$ & $(3.7 - 6.2) \!\times\! 10^{-5}$ & $(3 - 5)$\,MeV
\end{tabular}}
\label{tab:ZWmass}
\vspace{-4mm}
\end{table}
The situation changes when the precision on $M_W$ is significantly increased to be
comparable with the precision on $M_Z$. In \gtab{tab:ZWmass} we show the projected
$1\sigma$ precision of $Z$ and $W$ mass measurements at CEPC \cite{CEPC,Zpole}.
Comparing with the existing measurements in \gtab{tab:inputs2}, where the precision
on $M_W$ is almost an order of magnitude worse than the precision on $M_Z$,
the $M_W$ measurement at CEPC is relatively improved more than $M_Z$. Fixing $M_Z$ to
the experimental central value may disguise its interplay with $M_W$ as summarized
in \gtab{tab:ZWmass-effects}.
\begin{table}[h]
\vspace{-3mm}
\setlength{\tabcolsep}{0.8mm}
\centering
\tbl{Impacts of the projected $M_Z$ and $M_W$ measurements at CEPC \cite{CEPC,Zpole} on
         the reach of new physics scale $\Lambda/\sqrt{|c_j|}$ (in TeV) at 95\% C.L.
         The Higgs observables (including $\sigma(\nu \bar \nu h)$ at 350\,GeV)
         and the existing electroweak precision
         observables (Table \ref{tab:inputs2}) are always included in each row.
         The differences among the four rows arise from whether taking into account
         the measurements of $M_Z$ and $M_W$ (Table \ref{tab:ZWmass}) or not.
         The second (third) row contains the measurement of $M_Z$ ($M_W$) alone,
         while the first (last) row contains none (both) of them.
         We mark the entries of the most significant improvements from
         $M_Z$ and/or $M_W$ measurements in red color with an underscore.}
{
\begin{tabular}{cccccccccccccccc}
$\mathcal O_H^{}$ & $\mathcal O_T^{}$ & $\mathcal O_{WW}^{}$
& $\mathcal O_{BB}^{}$ & $\mathcal O_{WB}^{}$ & $\mathcal O_{HW}^{}$
& $\mathcal O_{HB}^{}$ & $\mathcal O^{(3)}_{LL}$ & $\mathcal O^{(3)}_L$
& $\mathcal O_L^{}$ & $\mathcal O_R^{}$ & $\mathcal O^{(3)}_{L,q}$
& $\mathcal O_{L,q}^{}$ & $\mathcal O_{R,u}^{}$ & $\mathcal O_{R,d}^{}$
& $\mathcal O_g^{}$ \\
\hline
 2.74 &       10.6  & 6.38 & 5.78 &       6.53  & 2.15 & 0.603 &       8.57  &       12.1  & 10.2 & 8.78 & 1.85 & 0.565 & 0.391 & 0.337 & 39.8 \\
 2.74 & \ured{10.7} & 6.38 & 5.78 & \ured{6.54} & 2.15 & 0.603 & \ured{8.61} &       12.1  & 10.2 & 8.78 & 1.85 & 0.565 & 0.391 & 0.337 & 39.8 \\
 2.74 & \ured{21.0} & 6.38 & 5.78 & \ured{10.4} & 2.15 & 0.603 & \ured{15.5} & \ured{16.4} & 10.2 & 8.78 & 1.85 & 0.565 & 0.391 & 0.337 & 39.8 \\
 2.74 & \ured{23.7} & 6.38 & 5.78 & \ured{11.6} & 2.15 & 0.603 & \ured{17.4} & \ured{18.1} & 10.2 & 8.78 & 1.85 & 0.565 & 0.391 & 0.337 & 39.8
\end{tabular}}
\label{tab:ZWmass-effects}
\vspace{-3mm}
\end{table}
On the basis of fitting Higgs and existing electroweak precision observables in
\gtab{tab:inputs2} (first row), adding better measurement on $M_Z$ (second row)
does not improve the reach on new physics scales while the $M_W$ measurement
at CEPC (third row) can significantly improve the results. The enhancement from
$M_W$ can be as large as a factor of 2 for ($\mathcal O_T$, $\mathcal O_{WB}$,
$\mathcal O^{(3)}_{LL}$). It is interesting to see that further imposing the
$M_Z$ measurement at CEPC, after already imposing $M_W$, the sensitivity can
be increased by another 10\%.

Another feature is the scale of $\mathcal O_H$ increases from 2.5\,TeV in
\gtab{tab:scheme} to 2.74\,TeV in \gtab{tab:ZWmass-effects} by 10\%. This is
because of the $WW$ fusion at $\sqrt s = 350$\,GeV which has been added into
the $\chi^2$ fit for \gtab{tab:ZWmass-effects}. From the Higgsstrahlung peak
at $\sqrt s = 250$\,GeV to $t\bar t$ threshould $\sqrt s = 350$\,GeV leads
to significant increase in $\sigma(\nu \bar \nu h)$ but decrease in
$\sigma(Zh)$. Consequently, we can expect the reduced uncertainty in
$\sigma(\nu \bar \nu h)$, estimated by TLEP \cite{FCC-ee} ,
is the major gain from increasing the lepton collider energy.
Nevertheless, the benifit for constraining new physics scales is just 10\%.

\subsection{The Improvement from $Z$-Pole Observables}

\begin{table}[h]
\centering
\tbl{Projected $Z$-pole measurements at CEPC \cite{CEPC,Zpole}
         with integrated luminosity of $5\,\mbox{ab}^{-1}$.}
{
% \begin{tabular}{cc}
% ~~~Observables~~~ & ~~~Relative Error~~~ \\
% \hline
% $N_\nu$           & $1.8 \times 10^{-3}$ \\
% $A_{FB}(b)$       & $1.5 \times 10^{-3}$ \\
% $R_b$             & $8 \times 10^{-4}$ \\
% $R_\mu$           & $5 \times 10^{-4}$ \\
% $R_\tau$          & $5 \times 10^{-4}$ \\
% $\sin^2 \theta_W$ & $1 \times 10^{-4}$
% \end{tabular}
\setlength{\tabcolsep}{3.8mm}
\setlength\extrarowheight{3pt}
\begin{tabular}{cccccc}
$N_\nu$              & $A_{FB}(b)$          & $R^b$              & $R^\mu$            & $R^\tau$           & $\sin^2 \theta_w$ \\
\hline
$1.8 \times 10^{-3}$ & $1.5 \times 10^{-3}$ & $8 \times 10^{-4}$ & $5 \times 10^{-4}$ & $5 \times 10^{-4}$ & $1 \times 10^{-4}$ \\
\end{tabular}
}
\label{tab:Zpole}
\end{table}
\begin{figure}[h]
\centering
\includegraphics[width=\textwidth]{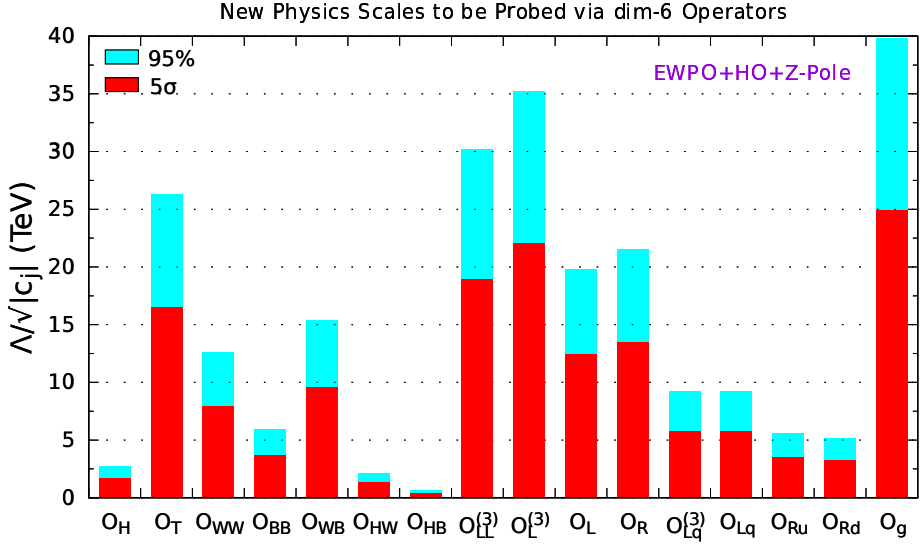}
\caption{The 95\% exclusion (blue) and $5\sigma$ discovery (red) sensitivities
to the new physics scales $\Lambda/\sqrt{|c_j|}$
by combining the current electroweak precision measurements ($\alpha$, $G_F$, $M_Z$, $M_W$) \cite{PDG14}
with the future Higgs observables at the Higgs factory CEPC (Table\,\ref{tab:inputs}) and $Z$-pole
measurements (Table\,\ref{tab:ZWmass}) under a projected luminosity of 5\,ab$^{-1}$ \cite{CEPC}.}
\label{fig:EW4}
\end{figure}

Finally, we further include the $Z$-pole measurements listed in \gtab{tab:Zpole}
and show the results in \gfig{fig:EW4}. Now the new physics scales can
reach as high as 35\,TeV, in addition to the 40\,TeV of $\mathcal O_g$.
Three operators ($\mathcal O^{(3)}_{LL}$, $\mathcal O^{(3)}_L$, $\mathcal O_g$)
can be probed up to above 30\,TeV. Among them, the scales of $\mathcal O^{(3)}_{LL}$
and $\mathcal O^{(3)}_L$ are enhanced by the $Z$-pole observables by almost another
factor of 2. Most of the operators shown in \gfig{fig:EW4} can be probed above
5\,TeV, including ($\mathcal O_T$, $\mathcal O_{WW}$, $\mathcal O_{BB}$,
$\mathcal O_{WB}$, $\mathcal O_L$, $\mathcal O_R$, $\mathcal O^{(3)}_{Lq}$,
$\mathcal O_{Lq}$, $\mathcal O_{Ru}$, $\mathcal O_{Rd}$).

Note that the most
significant improvement comes from quark related operators ($\mathcal O^{(3)}_{Lq}$,
$\mathcal O_{Lq}$, $\mathcal O_{Ru}$, $\mathcal O_{Rd}$). Roughly speaking, the
scale of $\mathcal O^{(3)}_{Lq}$ is increased by a factor of 5 while
($\mathcal O_{Lq}$, $\mathcal O_{Ru}$, $\mathcal O_{Rd}$) are increased by a
factor of 10. This is because the quark related operators cannot enter the
Higgs production cross sections $\sigma(Zh)$ or $\sigma(\nu \bar \nu h)$ but can
affect the $h \rightarrow ZZ$ and $h \rightarrow WW$ decays indirectly to some
extent. Since the Higgs boson mass at 125\,GeV is not large enough to put the
two $Z/W$ bosons on shell, the decay width has to be calculated with at least
one off-shell $Z/W$ boson. Actually, the contribution of two off-shell $Z/W$
can be as large as 25\%. Then, the decay width has to be evaluated for the
complete chains $h \rightarrow ZZ \rightarrow f_1 \bar f_i f_j \bar f_j$
($f$ denoting fermions) and $h \rightarrow WW \rightarrow u_i \bar d_j d_k \bar u_l$
($u$ for up-type fermion and $d$ for the down-type). The quark related operators
then enters as correction to the $Z f \bar f$ and $W^- u_i \bar d_j$ vertices.
Since the $h \rightarrow ZZ$ and $h \rightarrow WW$ channels do not have dominating
branching ratios, see \gtab{tab:inputs2}, and the precision on decay branching
ratios is not as good as the inclusive measurement of $\sigma(Zh)$, the new
physics scales probed by Higgs observables are very low. For ($\mathcal O_{L,q}$,
$\mathcal O_{R,u}$, $\mathcal O_{R,d}$), the sensitivity can only reach
$300 \sim 500$\,GeV while the scale for $\mathcal O^{(3)}_{L,q}$ is 1.85\,TeV.
With $Z$-pole measurements, where the $Z f \bar f$ and $W^- u_i \bar d_j$ vertices
dominate, the sensitivities are significantly enhanced to $5 \sim 9$\,TeV.

\section{Conclusion}
\label{sec:conclusion}

The discovery of the Higgs boson at LHC completes the particle spectrum of the SM.
Nevertheless, the SM as a whole can be claimed as complete only after fully testing
all the interactions that it dictates. The particle physics is not
just about particle but also interactions between them. Especially, the scalar
Higgs boson mediates new type of interactions. Our current understanding
of the Higgs coupling with fermions and Higgs self-interactions is not as good
as the gauge interactions. New physics may enter by modifying the Higgs coupling
with the SM particles. From the point of view of testing the SM or going beyond,
a lepton collider (Higgs factory) for precision measurement is necessary.

The next-generation lepton colliders, such as CEPC, FCC-ee, and ILC, are
motivated by the discovery of Higgs boson at LHC for precision measurement
of its properties. We present the extracted Higgs coupling precision from Higgs
observables at CEPC to show its physics potential. With one million events,
the Higgs couplings
can be measured to $\mathcal O(1\%)$ level. In particular, the Higgs coupling with
$Z$ boson can be as good as 0.25\% due to the inclusive cross section of
the Higgsstrahlung process $e^+ e^- \rightarrow Zh$.

Although CEPC runs at energy $\sqrt s = 250$\,GeV and $Z$-pole, it can
probe the new physics beyond the SM indirectly. We use dimension-6 operators
to estimate the new physics scales that can be reached at CEPC.
The result shows that the Higgs observables
together with existing electroweak precision observables can
constrain the new physics up to 10\,TeV at 95\% C.L. (40\,TeV for the gluon related
operator $\mathcal O_g$). If the $M_Z/M_W$ mass measurements and $Z$-pole
observables are also utilized, the new physics scale can be further pushed
up to 35\,TeV. The $Z$-pole observables are as important as the Higgs observables
and the $Z$-pole running is useful not only for the purpose of
calibration but also for probing the new physics scales. It is beneficial to
assign more time for $Z$-pole running at $\sqrt s \approx 90$\,GeV before
switching to the Higgs factory mode $\sqrt s = 250$\,GeV or returning to
$Z$-pole after finishing the Higgs observable measurements.
Since CEPC is a circular collider, the $50\sim100$\,km tunnel can host a
hadron ($pp$) collider SPPC with $\sqrt s = 50 \sim 100$\,TeV of energy after
CEPC. The energy scale ($10\sim 40$\,TeV) indirectly reachable at CEPC basically
covers the energy range to be effectively explored at SPPC. The lepton
collider CEPC can guide and pave the road for the sequent hadron collider SPPC.

Although our study takes CEPC for illustration, the conclusion can apply
to other candidate lepton colliders, such as FCC-ee and ILC.
The technical details for obtaining the results
presented here can be found in our formal paper \cite{dim6} .

\section{Acknowledgements}

We thank Matthew McCullough, Manqi Ruan, and Tevong You for 
many valuable discussions. We are grateful to Michael Peskin for discussing the analysis of Ref \cite{Peskin}. We also thank Timothy Barklow, Tao Han, Zhijun Liang and Matthew Strassler for 
discussions. SFG is grateful to the Jockey Club Institute for Advanced Study
at the Hong Kong University of Science and Technology, especially
Henry Tye and Tao Liu, for kind invitation to attend the IAS
Program/Conference on High Energy Physics and hospitality during
the stay.

\end{document}